# Static charged dilaton black hole cannot be overcharged by gedanken experiments

Jie Jiang,[2,*] Banglin Deng,[1,†] and Zhaohui Chen[2,‡]

[1]*Department of Applied Physics, College of Geophysics, Chengdu University of Technology, Chengdu 610059, Sichuan, China*
[2]*Department of Physics, Beijing Normal University, Beijing, 100875, China*

We consider the new version of the gedanken experiments proposed recently by Sorce and Wald to overcharge a static charged dilaton black hole. First of all, we derive the first-order and second-order perturbation inequalities in the Einstein-Maxwell-dilaton gravitational theory based on the Iyer-Wald formalism. As a result, we find that the weak cosmic censorship conjecture associated with this black hole can be protected after taking into account the second-order perturbation inequality, although violated by the scene without considering this inequality. Therefore, there is no violation of the weak cosmic censorship conjecture around the charged static dilaton black holes in Einstein-Maxwell-dilaton gravity.

## I. INTRODUCTION

When a singularity is not hidden behind a black hole horizon, so as to be seen by a distant observer, then it is called a naked singularity. And the singularity will violate the predictability of general relativity as a classical theory. Therefore, Penrose proposed the weak cosmic censorship conjecture (WCC), which asserts that singularities formed by the gravitational collapse of matter are hidden behind event horizons [1]. Even though there is still no general proof for this conjecture, many efforts have been taken for decades to test it [2]. Particularly, in a seminal work, Wald proposed a gedanken experiment to test this conjecture by examining whether the black hole horizon could be destroyed by plunging a test particle into a black hole [3]. The result shows that we cannot destroy an extremal Kerr-Newman (KN) black hole in this way. But as initiated by Hubeny in 1999 [4], the nearly extremal KN black hole can be destroyed by inputting the test particle [5–9]. And it also received lots of attention and followed by extensive studies in various theories [10–26].

Motivated by these results, Sorce and Wald [27] have recently suggested a new version of the gedanken experiment without proposing the test particle assumption of explicit analyses of trajectories of particle matter. In this version, they apply the Iyer-Wald formalism [28] as well as the null energy condition to the general matter perturbation on the black holes and obtain the first-order and second-order inequalities of collision matter. After the second-order perturbation inequality of the energy, angular momentum, and charge are taken into account, they showed that the nearly extremal Kerr-Newman black hole cannot be destroyed under the second-order approximation of the perturbation and no violation of the Hubeny type can ever occur.

Most recently, this new version has also been investigated in five-dimensional Myers-Perry black holes and higher-dimensional charged black holes, and they show that the WCC is well protected for the nearly extremal black holes when the second-order perturbation inequality is considered [29,30]. These black holes have a lot of remarkable properties; for example, all of them have two horizons. Therefore, it is natural for us to study whether the second-order perturbation inequality can ensure the WCC in all kinds of black holes, especially those with different causal structure. As one of the most interesting solutions of general relativity, the dilaton black hole has many different features from the above cases, where the inner horizon is taken placed by a singular surface after introducing the dilaton field (see Fig. 1). Therefore, its subextremal case shares the same causal structure as the Schwarzschild black hole. However, differing from the Schwarzschild case, as shown in Ref. [31], the static charged dilaton black hole in the Einstein-Maxwell-dilaton theory could be overcharged by the old version of the gedanken experiment, since the spacetime causal structure also relies on the electric charges. In this paper, we would like to consider the Hubeny scenario by this new version of gedanken experiment and investigate whether the WCC can be restored when the second-order correction is taken into consideration.

[*]jiejiang@mail.bnu.edu.cn
[†]Corresponding author.
bldeng@aliyun.com
[‡]chenzhaohui@mail.bnu.edu.cn

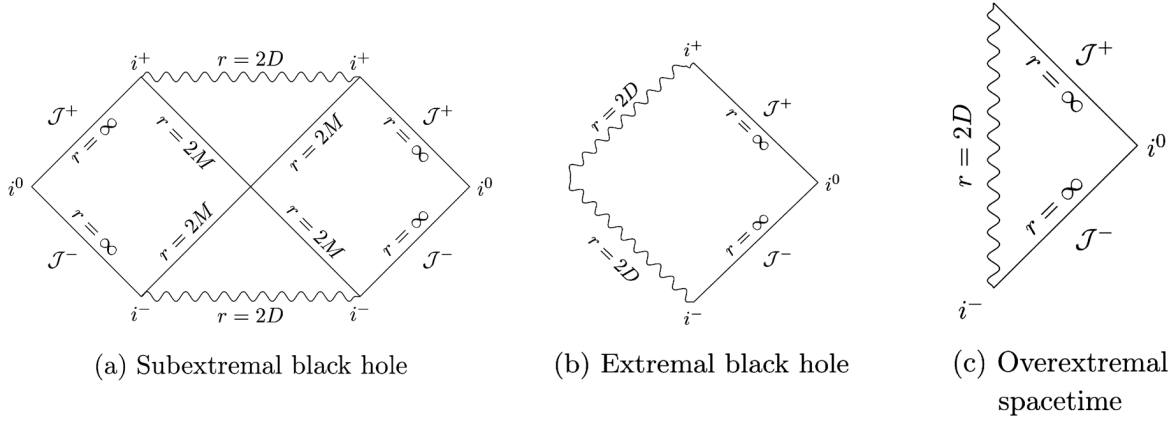

FIG. 1. Penrose diagrams representing the causal structure of the static charged dilaton blacks [32] for (a) the subextremal case, (b) the extremal case, and (c) the overextremal case.

Our paper is organized as follows. In the next section, we review the Iyer-Wald formalism for general diffeomorphism covariant theories and show the corresponding variation quantities. In Sec. III, we focus on dilaton black holes in the Einstein-Maxwell-dilaton theory and derive the relevant quantities in this case. In Sec. IV, we present the setup for the new version of the gedanken experiment and derive the first-order and second-order perturbation inequalities for the optimal first-order perturbation of the dilaton black holes. In Sec. V, we examine the Hubeny scenario from the new version of the gedanken experiment when the second-order perturbation inequality is considered and compare to the result without a second-order perturbation. Finally, conclusions are presented in Sec. VI.

## II. IYER-WALD FORMALISM AND VARIATIONAL IDENTITIES

In this paper, we would like to use the Noether charge formalism proposed by Iyer and Wald to investigate the gedanken experiments in the charged static black holes of Einstein-Maxwell-dilaton theory. First, we consider a general diffeomorphism-covariant theory on a four-dimensional spacetime $M$. The Lagrangian can be given by a 4-form $\boldsymbol{L}$ where the dynamical fields consist of a Lorentz signature metric $g_{ab}$ and other fields $\psi$. Following the notation in Ref. [28], we use boldface letters to denote differential forms and collectively refer to $(g_{ab}, \psi)$ as $\phi$. The variation of $\boldsymbol{L}$ with is given by

$$\delta \boldsymbol{L} = \boldsymbol{E}_\phi \delta \phi + d\boldsymbol{\Theta}(\phi, \delta\phi), \quad (1)$$

where $\boldsymbol{E}_\phi = 0$ gives the equations of motion of this theory and $\boldsymbol{\Theta}$ is called the symplectic potential 3-form which is locally constructed out of $\phi$, $\delta g_{ab}$, and their derivatives. The symplectic current 3-form is defined by

$$\boldsymbol{\omega}(\phi, \delta_1\phi, \delta_2\phi) = \delta_1\boldsymbol{\Theta}(\phi, \delta_2\phi) - \delta_2\boldsymbol{\Theta}(\phi, \delta_1\phi). \quad (2)$$

Let $\zeta^a$ be the infinitesimal generator of a diffeomorphism. By replacing $\delta$ by $\mathcal{L}_\zeta$ in (1), one can define the Noether current 3-form $\boldsymbol{J}_\zeta$ associated with $\zeta^a$:

$$\boldsymbol{J}_\zeta = \boldsymbol{\Theta}(\phi, \mathcal{L}_\zeta\phi) - \zeta \cdot \boldsymbol{L}. \quad (3)$$

A straightforward calculation yields

$$d\boldsymbol{J}_\zeta = -\boldsymbol{E}_\phi \mathcal{L}_\zeta\phi, \quad (4)$$

which indicates that $\boldsymbol{J}_\zeta$ is closed when the equations of motion are satisfied. On the other hand, it was shown in Ref. [33] that the Noether current can be written in the form

$$\boldsymbol{J}_\zeta = \boldsymbol{C}_\zeta + d\boldsymbol{Q}_\zeta, \quad (5)$$

where $\boldsymbol{Q}_\zeta$ is called the Noether charge and $\boldsymbol{C}_\zeta = \zeta^a \boldsymbol{C}_a$ are the constraints of the theory, i.e., $\boldsymbol{C}_a = 0$ when the equations of motion are satisfied.

Then, by keeping $\zeta^a$ fixed and comparing the variations of (3) and (5), one can obtain the first variational identity

$$d[\delta\boldsymbol{Q}_\zeta - \zeta \cdot \boldsymbol{\Theta}(\phi, \delta\phi)] = \boldsymbol{\omega}(\phi, \delta\phi, \mathcal{L}_\zeta\phi) - \zeta \cdot \boldsymbol{E}\delta\phi - \delta\boldsymbol{C}_\zeta. \quad (6)$$

The second variational identity can further be obtained, and it can be shown as

$$d[\delta^2\boldsymbol{Q}_\zeta - \zeta \cdot \delta\boldsymbol{\Theta}(\phi, \delta\phi)] = \boldsymbol{\omega}(\phi, \delta\phi, \mathcal{L}_\zeta\delta\phi) \\ - \zeta \cdot \delta\boldsymbol{E}\delta\phi - \delta^2\boldsymbol{C}_\zeta, \quad (7)$$

where we also used the equations of motion and assume $\zeta^a$ is a symmetry of $\phi$, i.e., $\mathcal{L}_\zeta \phi = 0$. In what follows, we shall consider the globally hyperbolic static solution with a timelike Killing vector $\xi^a$ such that $\mathcal{L}_\xi \phi = 0$. The Arnowitt-Deser-Misner mass of this black hole is given by

$$\delta M = \int_\infty [\delta \boldsymbol{Q}_\xi - \xi \cdot \boldsymbol{\Theta}(\phi, \delta\phi)]. \tag{8}$$

Supposing that $\Sigma$ is a hypersurface with a cross section $B$ of the horizon and the spatial infinity as its boundaries, the integration of the first and second variational identities can be written as

$$\delta M = \int_B [\delta \boldsymbol{Q}_\xi - \xi \cdot \boldsymbol{Q}(\phi, \delta\phi)] - \int_\Sigma \delta \boldsymbol{C}_\xi,$$
$$\delta^2 M = \int_B [\delta^2 \boldsymbol{Q}_\xi - \xi \cdot \delta\boldsymbol{Q}(\phi, \delta\phi)]$$
$$- \int_\Sigma \xi \cdot \delta\boldsymbol{E}\delta\phi - \int_\Sigma \delta\boldsymbol{C}_\xi + \mathcal{E}_\Sigma(\phi, \delta\phi), \tag{9}$$

where we denote

$$\mathcal{E}_\Sigma(\phi, \delta\phi) = \int_\Sigma \boldsymbol{\omega}(\phi, \delta\phi, \mathcal{L}_\xi \delta\phi). \tag{10}$$

## III. EINSTEIN-MAXWELL-DILATON THEORY AND ITS STATIC SOLUTION

For our purpose, in this section, we consider an Einstein-Maxwell-dilaton theory in four-dimensional spacetime with the following Lagrangian:

$$\boldsymbol{L} = \frac{1}{16\pi}(R - 2\nabla_a \psi \nabla^a \psi - e^{-2\psi}\mathcal{F})\boldsymbol{\epsilon}, \tag{11}$$

where $\mathcal{F} = F_{ab}F^{ab}$ with the electromagnetic strength $\boldsymbol{F} = d\boldsymbol{A}$. This model describes a massless dilaton scalar field coupled to the linear electromagnetic field. The symplectic potential can be given by

$$\boldsymbol{\Theta}(\phi, \delta\phi) = \boldsymbol{\Theta}^{\mathrm{GR}}(\phi, \delta\phi) + \boldsymbol{\Theta}^{\mathrm{EM}}(\phi, \delta\phi) + \boldsymbol{\Theta}^{\mathrm{DIL}}(\phi, \delta\phi) \tag{12}$$

with

$$\boldsymbol{\Theta}^{\mathrm{GR}}_{abc}(\phi, \delta\phi) = \frac{1}{16\pi}\epsilon_{dabc}g^{de}g^{fg}(\nabla_g \delta g_{ef} - \nabla_e \delta g_{fg}),$$
$$\boldsymbol{\Theta}^{\mathrm{EM}}_{abc}(\phi, \delta\phi) = -\frac{1}{4\pi}\epsilon_{dabc}G^{de}\delta A_e,$$
$$\boldsymbol{\Theta}^{\mathrm{DIL}}_{abc}(\phi, \delta\phi) = -\frac{1}{4\pi}\epsilon_{dabc}(\nabla^d \psi)\delta\psi. \tag{13}$$

Here we have defined

$$\boldsymbol{G} = e^{-2\psi}\boldsymbol{F}. \tag{14}$$

The Noether charge is given by

$$\boldsymbol{Q}_\xi = \boldsymbol{Q}^{\mathrm{GR}}_\xi + \boldsymbol{Q}^{\mathrm{EM}}_\xi, \tag{15}$$

where

$$(\boldsymbol{Q}^{\mathrm{GR}}_\xi)_{ab} = -\frac{1}{16\pi}\epsilon_{abcd}\nabla^c \xi^d,$$
$$(\boldsymbol{Q}^{\mathrm{EM}}_\xi)_{ab} = -\frac{1}{8\pi}\epsilon_{abcd}G^{cd}A_e\xi^e. \tag{16}$$

If the additional charged matter sources are taken into account, the equations of motion can be written as

$$R_{ab} - \frac{1}{2}Rg_{ab} = 8\pi(T^{\mathrm{EM}}_{ab} + T^{\mathrm{DIL}}_{ab} + T_{ab}),$$
$$\nabla_a G^{ab} = 4\pi j^a,$$
$$E_\psi = \frac{1}{4\pi}\left(\nabla^2 \psi + \frac{1}{2}e^{-2\psi}\mathcal{F}\right) = 0, \tag{17}$$

with

$$T^{\mathrm{EM}}_{ab} = \frac{e^{-2\psi}}{4\pi}\left(F_{ac}F_b{}^c - \frac{1}{4}g_{ab}\mathcal{F}\right),$$
$$T^{\mathrm{DIL}}_{ab} = \frac{1}{4\pi}\left(\nabla_a \psi \nabla_b \psi - \frac{1}{2}g_{ab}\nabla_c \psi \nabla^c \psi\right). \tag{18}$$

Here $T_{ab}$ corresponds to the nonelectromagnetic and dilaton part of the stress-energy tensor, $j^a$ corresponds to the electromagnetic charge current, and both of them are nonvanishing after the matter source is introduced. Then, the equations of motion part and constraints in Eq. (1) for the Einstein-Maxwell-dilaton theory are given by

$$\boldsymbol{E}(\phi)\delta\phi = -\boldsymbol{\epsilon}\left[\frac{1}{2}T^{ab}\delta g_{ab} + j^a \delta A_a + E_\psi \delta\psi\right],$$
$$\boldsymbol{C}_{abcd} = \epsilon_{ebcd}(T_a{}^e + A_a j^e). \tag{19}$$

If the background spacetime is stationary, the flux of the stress-energy tensors of the electromagnetic field and dilaton field through the horizon must vanish. From (18), it implies that $F^{ab}$ must take the form

$$F^{ab} = v^{[a}k^{b]} + w^{ab} \tag{20}$$

and $\psi$ satisfies $\mathcal{L}_\xi \psi = 0$, where $w^{ab}$ is a purely tangential to the horizon [27].

From (13), the symplectic current for the Einstein-Maxwell-dilaton theory can be written as

$$\boldsymbol{\omega}(\phi, \delta_1\phi, \delta_2\phi) = \boldsymbol{\omega}^{\mathrm{GR}}_{abc} + \boldsymbol{\omega}^{\mathrm{EM}}_{abc} + \boldsymbol{\omega}^{\mathrm{DIL}}_{abc}, \tag{21}$$

where

$$\omega_{abc}^{\text{GR}} = \frac{1}{16\pi}\epsilon_{dabc}w^d,$$

$$\omega_{abc}^{\text{EM}} = \frac{1}{4\pi}[\delta_2(\epsilon_{dabc}G^{de})\delta_1 A_e - \delta_1(\epsilon_{dabc}G^{de})\delta_2 A_e],$$

$$\omega_{abc}^{\text{DIL}} = \frac{1}{4\pi}[\delta_2(\epsilon_{dabc}\nabla^d\psi)\delta_1\psi - \delta_1(\epsilon_{dabc}\nabla^d\psi)\delta_2\psi], \quad (22)$$

in which we denote

$$w^a = P^{abcdef}(\delta_2 g_{bc}\nabla_d\delta g_{ef} - \delta_1 g_{bc}\nabla_d\delta_2 g_{ef}) \quad (23)$$

with

$$P^{abcdef} = g^{ae}g^{fb}g^{cd} - \frac{1}{2}g^{ad}g^{be}g^{fc} - \frac{1}{2}g^{ab}g^{cd}g^{ef}$$
$$- \frac{1}{2}g^{bc}g^{ae}g^{fd} + \frac{1}{2}g^{bc}g^{ad}g^{ef}. \quad (24)$$

We next restrict on the charged static spherically symmetric solution of the four-dimensional Einstein-Maxwell-Dilaton theory, which can be described by [32]

$$ds^2 = -\left(1 - \frac{2M}{r}\right)dt^2 + \left(1 - \frac{2M}{r}\right)^{-1}dr^2$$
$$+ r^2\left(1 - \frac{2D}{r}\right)(d\theta^2 + \sin^2\theta d\varphi^2),$$
$$\mathbf{F} = -\frac{Q}{r^2}dt \wedge dr, \qquad e^{2\psi} = 1 - \frac{2D}{r}, \quad (25)$$

with the constraint

$$Q^2 = 2MD. \quad (26)$$

It is also known as the Gibbons-Maeda-Garfinke-Horowitz-Stromninger solution. The Penrose diagrams of this solution are shown in Fig. 1. This black hole solution exists as long as censorship condition $M > D$ is satisfied. And the horizon is located at $r_h = 2M$. We shall refer the extremal limit to the case $Q^2 = 2M^2$, where the singular radius $r = 2D$ coincides with the horizon. Then, the area of the event horizon with the area is given by

$$A = 8\pi(2M^2 - Q^2). \quad (27)$$

One can note that our event horizon is also a Killing horizon which is generated by the Killing field $\xi^a = (\partial/\partial t)^a$. And the corresponding horizon electric potential and surface gravity can be read off

$$\Phi_{\mathcal{H}} = -\xi^a A_a|_{r=2M} = \frac{Q}{2M},$$
$$\kappa = \frac{1}{2}f'(r_h) = \frac{1}{4M}. \quad (28)$$

## IV. PERTURBATION INEQUALITIES OF GEDANKEN EXPERIMENTS

As in the new gedanken experiment designed in Ref. [27], the situation we plan to investigate is what happens to the above static dilaton black holes when they are perturbed by a one-parameter family of the matter source according to Einstein equation as well as the Maxwell equation

$$G_{ab}(\lambda) = 8\pi[T_{ab}^{\text{EM}}(\lambda) + T_{ab}(\lambda)],$$
$$\nabla_a^{(\lambda)}[e^{-2\psi(\lambda)}F^{ab}(\lambda)] = 4\pi j^a(\lambda), \quad (29)$$

around $\lambda = 0$ with $T^{ab}(0) = 0$ and $j^a[0] = 0$. Without loss of generality, we shall assume all the matter goes into the black hole through a finite portion of the future horizon; i.e., the matter source $\delta T^{ab}$ and $\delta j^a$ are nonvanishing only in a compact region of the future horizon. In order to obtain the first-order and second-order perturbations of the black hole, with the similar consideration of Ref. [27], we also introduce the following assumption.

*Additional assumption.*—The nonextremal, unperturbed static charged dilaton black hole is linearly stable to perturbations; i.e., any source-free solution to the linearized Einstein-Maxwell-dilaton equations approaches a perturbation towards another static charged dilaton black hole at sufficiently late times.

With these in mind, we can always choose a hypersurface $\Sigma = \mathcal{H} \cup \Sigma_1$ such that it starts from the bifurcate surface $B$ of the unperturbed horizon, continues up the horizon through the portion $\mathcal{H}$ till the very late cross section $B_1$ where the matter source vanishes, and then becomes spacelike as $\Sigma_1$ to approach the spatial infinity. By considering the additional assumption, the dynamical fields satisfy the source-free equation of motion $\mathbf{E}[\phi(\lambda)] = 0$ on the portion $\Sigma_1$, and the solution is described by Eq. (25).

Then, if we work with the Gaussian null coordinates near the unperturbed horizon, we can further obtain

$$\int_B \mathbf{Q}_\xi(\lambda) = \frac{\kappa}{8\pi}A_B(\lambda) \quad (30)$$

with $A_B$ the area of the bifurcate surface [3]. With the above preparation, we now derive the first-order inequality obeyed by the perturbation at $\lambda = 0$. Note that for our choice the perturbation vanishes on the bifurcate surface $B$ and the first equation of Eq. (9) reduces to

$$\delta M = -\int_\Sigma \delta C_\xi = -\int_\mathcal{H} \epsilon_{ebcd}[\delta T_a{}^e + A_a \delta j^e]\xi^a, \quad (31)$$

where we used the fact that $T^{ab} = j^a = 0$ in the background spacetime. Since $\Phi = -\xi^a A_a$ is constant on $\mathcal{H}$, we may pull it out of the integral. The integral $\delta Q_{\text{flux}} = \int_\mathcal{H} \delta(\epsilon_{ebcd} j^e)$ is just the total flux of electromagnetic charge through the horizon. Since all of the charge added to the spacetime falls through the horizon, this flux is just equal to the total perturbed charge of the black hole, $\delta Q_{\text{flux}} = \delta Q$. Combining these observations yields the following formula relating the perturbed parameters of the black hole spacetime:

$$\delta M - \Phi_H \delta Q = -\int_\mathcal{H} \epsilon_{ebcd} \delta T_a{}^e \xi^a = \int_\mathcal{H} \tilde{\epsilon}\delta T_{ab} k^a \xi^b, \quad (32)$$

where $\tilde{\epsilon}$ is the corresponding volume element on the horizon, which is defined by $\epsilon_{ebcd} = -4k_{[e}\tilde{\epsilon}_{bcd]}$ with the future-directed normal vector $k^a \propto \xi^a$ on the horizon. Then, according to the null energy condition $\delta T_{ab} k^a k^b \geq 0$, (32) yields the inequality

$$\delta M - \Phi_H \delta Q \geq 0. \quad (33)$$

Obviously, if we want to violate $M^2 - 2Q^2 \geq 0$, the optimal choice is to saturate (33) by requiring $\delta T_{ab} k^a k^b|_\mathcal{H} = 0$; namely, i.e., the energy flux through the horizon vanished for the first-order nonelectromagnetic perturbation. Then, (32) becomes

$$\delta M - \Phi_H \delta Q = 0. \quad (34)$$

The first-order perturbation of the Raychaudhuri equation

$$\frac{d\vartheta(\lambda)}{du} = -\frac{1}{3}\vartheta(\lambda)^2 - \sigma_{ab}(\lambda)\sigma^{ab}(\lambda) - R_{ab}(\lambda)k^a k^b \quad (35)$$

implies that $\delta\vartheta = 0$ on the horizon if we choose a gauge in which the first-order perturbed horizon coincides with the unperturbed one. Next, we consider the second-order inequality. By performing a similar analysis to the first-order result, the second equation of (9) reduces to

$$\delta^2 M = -\int_\mathcal{H} \xi \cdot \delta E \delta\phi - \int_\mathcal{H} \delta C_\xi + \mathcal{E}_\Sigma(\phi, \delta\phi). \quad (36)$$

Here, the integrals in the last two terms depend only on the surface $\mathcal{H}$ because $\delta E$ and $\delta^2 C_\xi$ vanishes on $\Sigma_1$ by the assumption that there is no source outside the black hole at late times. Moreover, since $\xi^a$ is tangent to the horizon, the first term vanishes. For the second term, together with (19), we have

$$(\delta^2 C_\xi)_{abc} = \delta^2(\epsilon_{eabc} T_d{}^e \xi^d) + \delta^2(\epsilon_{eabc} A_d j^e \xi^d). \quad (37)$$

Following the setting of Ref. [27], here we also impose the condition $\xi^a \delta A_a|_\mathcal{H} = 0$ by a gauge transformation, and we have

$$\delta^2\left[\int_\mathcal{H} \xi^a A_a \epsilon_{ebcd} j^e\right] = -\Phi_H \delta^2\left[\int_\mathcal{H} \epsilon_{ebcd} j^e\right]$$
$$= -\Phi_h \delta^2 Q_{\text{flux}} = -\Phi_h \delta^2 Q, \quad (38)$$

where $\delta^2 Q$ is the second-order change in charge of the black hole. Furthermore, by using the assumption that the first-order perturbation is optimal, we have

$$\delta^2 M - \Phi_H \delta^2 Q = \mathcal{E}_\Sigma(\phi, \delta\phi) - \int_\mathcal{H} \xi^a \epsilon_{ebcd} \delta^2 T_a{}^e$$
$$= \mathcal{E}_\Sigma(\phi, \delta\phi) + \int_\mathcal{H} \tilde{\epsilon}\delta^2 T_{ab} \xi^a k^b$$
$$\geq \mathcal{E}_\mathcal{H}(\phi, \delta\phi) + \mathcal{E}_{\Sigma_1}(\phi, \delta\phi), \quad (39)$$

where we have used the energy condition for the second-order perturbed nonelectromagnetic stress-energy tensor in the last step.

Next, we turn to compute the horizon contribution. It can be decomposed into

$$\mathcal{E}_\mathcal{H}(\phi, \delta\phi) = \int_\mathcal{H} \omega^{\text{GR}} + \int_\mathcal{H} \omega^{\text{EM}} + \int_\mathcal{H} \omega^{\text{DIL}}. \quad (40)$$

From the calculation in Ref. [27], the gravitational contribution is given by

$$\int_\mathcal{H} \omega^{\text{GR}} = \frac{1}{4\pi}\int_\mathcal{H} (\xi^a \nabla_a u)\delta\sigma_{ac}\delta\sigma^{bc}\tilde{\epsilon} \geq 0. \quad (41)$$

Then, we calculate the contribution for the electromagnetic part. From (22), we have

$$\omega^{\text{EM}}_{abc} = \frac{1}{4\pi}\epsilon_{dabc}[\delta A_e \mathcal{L}_\xi \delta G^{de} - \delta G^{de} \mathcal{L}_\xi \delta A_e]$$
$$+ \frac{1}{4\pi}[(\mathcal{L}_\xi \delta\epsilon_{dabc})G^{de}\delta A_e - \delta\epsilon_{dabc} G^{de} \mathcal{L}_\xi \delta A_e]. \quad (42)$$

By considering the gauge condition $\xi^a \delta A_a = 0$ on the horizon as well as (21), the last two terms will vanish. Then, Eq. (42) can be written as

$$\omega^{\text{EM}}_{abc} = \frac{1}{4\pi}\mathcal{L}_\xi(\epsilon_{dabc}\delta A_e \delta G^{de}) - \frac{1}{2\pi}\epsilon_{dabc}\delta G^{de}\mathcal{L}_\xi \delta A_e. \quad (43)$$

By considering

$$\mathcal{L}_\xi \eta = d(\xi \cdot \eta) \quad (44)$$

on the horizon, the integral over $\mathcal{H}$ of the first term on the right side will contribute only a boundary term at $S = \mathcal{H} \cap \Sigma_1$. With the fact that the perturbation is stationary at $S$, i.e., $\delta G_{ab}$ has the form (21). Together with the gauge condition $\xi^a \delta A_a = 0$ on $\mathcal{H}$, the first term of (43) makes no contribution to (42). Combining the above results, we have

$$\int_{\mathcal{H}} \omega_{abc}^{\text{EM}} = -\frac{1}{2\pi} \int_{\mathcal{H}} \epsilon_{dabc} \delta G^{de} \mathcal{L}_\xi \delta A_e$$
$$= -\frac{1}{2\pi} \int_{\mathcal{H}} \epsilon_{dabc} \xi^e \delta G^{de} \delta F_{fe}$$
$$= \frac{1}{2\pi} \int_{\mathcal{H}} \tilde{\epsilon}_{abc} k_d \xi^e \delta G^{de} \delta F_{fe}. \quad (45)$$

Finally, we evaluate the dilaton contribution. From (22), we have

$$\omega_{abc}^{\text{DIL}} = \frac{1}{4\pi} \epsilon_{dabc} [\mathcal{L}_\xi (\nabla^d \delta \psi) \delta \psi - (\nabla^d \delta \psi) \mathcal{L}_\xi \delta \psi]$$
$$+ \frac{1}{4\pi} [\mathcal{L}_\xi \delta \epsilon_{dabc} (\nabla^d \psi) \delta \psi - \delta \epsilon_{dabc} (\nabla^d \psi) \mathcal{L}_\xi \delta \psi]. \quad (46)$$

When pulled back to $\mathcal{H}$, the index $d$ must contribute a $k_d \propto \xi_d$. Then, since the background field is stationary, i.e., $\mathcal{L}_\xi \psi = \xi_d \nabla^d \psi = 0$, the last two terms vanish. Equation (46) becomes

$$\omega_{abc}^{\text{DIL}} = \frac{1}{4\pi} \mathcal{L}_\xi [\epsilon_{dabc} (\nabla^d \delta \psi) \delta \psi] - \frac{1}{2\pi} \epsilon_{dabc} (\nabla^d \delta \psi) \mathcal{L}_\xi \delta \psi. \quad (47)$$

With similar analysis as (43), the integral over $\mathcal{H}$ of the first term in (47) contributes only a boundary term at $S$. By considering $\delta \psi$ is stationary, i.e., $\mathcal{L}_\xi \delta \psi = 0$, this term also makes no contribution. Then, we have

$$\int_{\mathcal{H}} \omega^{\text{DIL}} = \frac{1}{2\pi} \int_{\mathcal{H}} \tilde{\epsilon} \xi^e k^d \nabla_e \delta \psi \nabla_d \delta \psi. \quad (48)$$

Together with (45), we have

$$\mathcal{E}_{\mathcal{H}}(\phi, \delta \phi) = \int_{\mathcal{H}} \tilde{\epsilon} \xi^a k^b (\delta^2 T_{ab}^{\text{EM}} + \delta^2 T_{ab}^{\text{DIL}}) \geq 0, \quad (49)$$

where we have used the null energy condition for the electromagnetic and dilaton stress-energy tensors. Finally, (39) reduces to

$$\delta^2 M - \Phi_H \delta^2 Q \geq \mathcal{E}_{\Sigma_1}(\phi, \delta \phi). \quad (50)$$

Now we are left to evaluate $\mathcal{E}_{\Sigma_1}(\phi, \delta \phi)$. To calculate it, we follow the trick introduced in Ref. [27] and write $\mathcal{E}_{\Sigma_1}(\phi, \delta \phi) = \mathcal{E}_{\Sigma_1}(\phi, \delta \phi^{\text{DL}})$, where $\phi^{\text{DL}}$ is introduced by the variation of a family of dilaton black hole solutions (25),

$$M^{\text{DL}}(\lambda) = M + \lambda \delta M, \qquad Q^{\text{DL}}(\lambda) = Q + \lambda \delta Q, \quad (51)$$

where $\delta M$ and $\delta Q$ are chosen to be in agreement with the first-order variation of the above optimal perturbation by the matter source. From the variation (51), one can find $\delta^2 M = \delta E = \delta^2 C = \mathcal{E}_{\mathcal{H}}(\phi, \delta \phi^{\text{DL}}) = 0$. Thus, from the second expression of (9), we have

$$\mathcal{E}_{\Sigma_1}(\phi, \delta \phi^{\text{DL}}) = -\int_B [\delta^2 Q_\xi - \xi \cdot \delta \Theta(\phi, \phi^{\text{DL}})]. \quad (52)$$

Since $\xi^a = 0$ on the bifurcation surface $B$, it can be expressed as

$$\mathcal{E}_{\Sigma_1}(\phi, \delta \phi^{\text{DL}}) = -\frac{\kappa}{8\pi} \delta^2 A_B^{\text{DL}}. \quad (53)$$

Therefore, the second-order inequality becomes

$$\delta^2 M - \Phi_H \delta^2 Q \geq -\frac{\kappa}{8\pi} \delta^2 A_B^{\text{DL}}. \quad (54)$$

The right sight of this inequality can be evaluated by taking two variations of the area formula $A_B = 8\pi(2M^2 - Q^2)$ and is given by

$$\delta^2 A_B^{\text{DL}} = 16\pi(2\delta M^2 - \delta Q^2). \quad (55)$$

Together with the optimal first-order inequality, the second-order inequality becomes

$$\delta^2 M - \Phi_H \delta^2 Q \geq \frac{(2M^2 - Q^2)\delta Q^2}{4M^3}. \quad (56)$$

## V. GEDANKEN EXPERIMENTS TO DESTROY A NEARLY EXTREMAL DILATON BLACK HOLE

In this section, we will explore the gedanken experiments to overcharge a nonextremal black hole by the physical process described above. Therefore, we define a function of $\lambda$ as

$$h(\lambda) = 2M(\lambda)^2 - Q(\lambda)^2. \quad (57)$$

Under the second-order approximation of $\lambda$, we have

$$h(\lambda) = (2M^2 - Q^2) + 4M\left(\delta M - \frac{Q}{2M}\delta Q\right)\lambda$$
$$+ 2M\left(\delta^2 M - \frac{Q}{2M}\delta^2 Q + \frac{1}{M}\delta M^2 - \frac{1}{2M}\delta Q^2\right)\lambda^2. \quad (58)$$

First, we would like to analyze the result found in Ref. [31] for the old version of gedanken experiments, where they consider only the perturbation of the test particle.

Therefore, in their case, there exists only a linear variation of the mass and charge of this black hole, and the second-order variation of black hole mass and charge vanish, i.e.,

$$M(\lambda) = M + \lambda \delta M, \qquad Q(\lambda) = Q + \lambda \delta Q. \quad (59)$$

Then, we have

$$h(\lambda) = (2M^2 - Q^2) + 4M\left(\delta M - \frac{Q}{2M}\delta Q\right)\lambda$$
$$+ 2M\left(\frac{1}{M}\delta M^2 - \frac{1}{2M}\delta Q^2\right)\lambda^2. \quad (60)$$

By using the optimal first-order inequality, it becomes

$$h(\lambda) = \frac{(2M^2 - Q^2)(2M^2 - \lambda^2 \delta Q^2)}{2M}. \quad (61)$$

According to this equation, we can see that, if we impose that the mass and charge of the background black hole have the same order as $\lambda$, then we can note that it is possible to make $h(\lambda) < 0$ for the nonextremal black holes, suggesting that the black hole could be overcharged if we neglect the second-order variation of mass and charge.

Next, we consider the new version of the gedanken experiments. Using the first-order and second-order inequalities (34) and (56), under the second-order approximation of the perturbation, we can further obtain

$$h(\lambda) \geq 2M^2 - Q^2 > 0, \quad (62)$$

where we have used the fact that the background spacetime has the black hole geometry. Thus, as we can see, when the second-order correction of the perturbation is taken into account, this static dilaton black hole cannot be overcharged.

## VI. CONCLUSION

It is shown in Ref. [31] that the old version of the gedanken experiment can destroy the static dilaton black holes in Einstein-Maxwell-dilaton theory if the backreaction or self-energy is ignored. However, in this paper, following a similar consideration as in Ref. [27], we showed that, after the second-order perturbation inequality is taken into account, the charged static dilaton black hole cannot be overcharged. Therefore, there is no violation of the weak cosmic censorship conjecture around the charged static dilaton black holes in Einstein-Maxwell-dilaton gravity. This result might indicate that, once this black hole is formed, it will never be overspun classically. Moreover, the above results indicate that the second-order perturbation inequality might play the role of the backreaction or self-energy for collision matter.

## ACKNOWLEDGMENTS

This research was supported by National Natural Science Foundation of China (NSFC) with Grants No. 11375026 and No. 11675015, the Cultivating Program of Excellent Innovation Team of Chengdu University of Technology (Grant No. KYTD201704), the Cultivating Program of Middle-aged Backbone Teachers of Chengdu University of Technology (Grant No. 10912-2019KYGG01511), and the Open Research Fund of Computational Physics Key Laboratory of Sichuan Province, Yibin University (Grant No. JSWL2018KFZ01).

[1] R. Penrose, Gen. Relativ. Gravit. **34**, 1141 (2002).
[2] R. M. Wald, arXiv:gr-qc/9710068.
[3] R. M. Wald, Ann. Phys. (N.Y.) **82**, 548 (1974).
[4] V. E. Hubeny, Phys. Rev. D **59**, 064013 (1999).
[5] F. de Felice and Y. Yu, Classical Quantum Gravity **18**, 1235 (2001).
[6] S. Hod, Phys. Rev. D **66**, 024016 (2002).
[7] T. Jacobson and T. P. Sotiriou, Phys. Rev. Lett. **103**, 141101 (2009).
[8] G. Chirco, S. Liberati, and T. P. Sotiriou, Phys. Rev. D **82**, 104015 (2010).
[9] A. Saa and R. Santarelli, Phys. Rev. D **84**, 027501 (2011).
[10] T. Jacobson and T. P. Sotiriou, Phys. Rev. Lett. **103**, 141101 (2009).
[11] T. Jacobson and T. P. Sotiriou, J. Phys. Conf. Ser. **222**, 012041 (2010).
[12] G. E. A. Matsas and A. R. R. da Silva, Phys. Rev. Lett. **99**, 181301 (2007).
[13] A. Saa and R. Santarelli, Phys. Rev. D **84**, 027501 (2011).
[14] S. Hod, Phys. Rev. D **66**, 024016 (2002).
[15] S. Gao and Y. Zhang, Phys. Rev. D **87**, 044028 (2013).
[16] Z. Li and C. Bambi, Phys. Rev. D **87**, 124022 (2013).
[17] K. Dzta and I. Semiz, Phys. Rev. D **88**, 064043 (2013).
[18] K. Dzta, Gen. Relativ. Gravit. **46**, 1709 (2014).
[19] G. Z. Tth, Classical Quantum Gravity **33**, 115012 (2016).
[20] B. Gwak and B. H. Lee, J. Cosmol. Astropart. Phys. 02 (2016) 015.
[21] B. Gwak and B. H. Lee, Phys. Lett. B **755**, 324 (2016).
[22] V. Cardoso and L. Queimada, Gen. Relativ. Gravit. **47**, 150 (2015).
[23] K. S. Revelar and I. Vega, Phys. Rev. D **96**, 064010 (2017).
[24] J. Sorce and R. M. Wald, Phys. Rev. D **96**, 104014 (2017).

[25] G. Chirco, S. Liberati, and T. P. Sotiriou, Phys. Rev. D **82**, 104015 (2010).
[26] H. M. Siahaan, Phys. Rev. D **93**, 064028 (2016).
[27] J. Sorce and R. M. Wald, Phys. Rev. D **96**, 104014 (2017).
[28] V. Iyer and R. M. Wald, Phys. Rev. D **50**, 846 (1994).
[29] J. An, J. Shan, H. Zhang, and S. Zhao, Phys. Rev. D **97**, 104007 (2018).
[30] B. Ge, Y. Mo, S. Zhao, and J. Zheng, Phys. Lett. B **783**, 440 (2018).
[31] T. Y. Yu and W. Y. Wen, Phys. Lett. B **781**, 713 (2018).
[32] P. Aniceto and J. V. Rocha, J. High Energy Phys. 05 (2017) 035.
[33] V. Iyer and R. M. Wald, Phys. Rev. D **52**, 4430 (1995).